\documentclass[aps,prb,twocolumn,superscriptaddress,showpacs]{revtex4}

\usepackage{amsmath}
\usepackage{amssymb}

\usepackage{graphicx}

\begin{document}

\title{Condon Domain Phase Diagram and Hysteresis Size for Beryllium}

\author{R.B.G.~Kramer}
\affiliation{LNCMI, CNRS, BP 166, 38042 Grenoble Cedex 9, France}
\affiliation{Max-Planck-Institut f\"{u}r Festk\"{o}rperforschung, Heisenbergstra{\ss}e 1, 70569 Stuttgart, Germany}
\affiliation{Institut N\'eel, CNRS--Universit\'e Joseph Fourier, BP 166, 38042 Grenoble Cedex 9, France}

\author{V.S.~Egorov}
\affiliation{LNCMI, CNRS, BP 166, 38042 Grenoble Cedex 9, France}
\affiliation{Max-Planck-Institut f\"{u}r Festk\"{o}rperforschung, Heisenbergstra{\ss}e 1, 70569 Stuttgart, Germany}
\affiliation{Russian Research Center "Kurchatov Institute", 123182 Moscow, Russia}

\author{A.G.M.~Jansen}
\affiliation{Service de Physique Statistique, Magn\'{e}tisme, et
Supraconductivit\'{e}, INAC, CEA-Grenoble, F-38054 Grenoble Cedex 9, France}

\author{W. Joss}
\affiliation{LNCMI, CNRS, BP 166, 38042 Grenoble Cedex 9, France}
\affiliation{Max-Planck-Institut f\"{u}r Festk\"{o}rperforschung, Heisenbergstra{\ss}e 1, 70569 Stuttgart, Germany}
\affiliation{Universit\'e Joseph Fourier, BP 53, 38041 Grenoble Cedex 9, France}

\begin{abstract}
The Condon domain phase diagram for beryllium is determined in magnetic fields up to 10~T
and at temperatures down to 1.3~K using a standard ac pick-up coil method to measure the de Haas-van
Alphen (dHvA) effect. The detection of the transition
point from the homogeneous state to the Condon domain state (CDS)
is based on the extremely non-linear response to the modulation
field resulting from a small irreversibility in the dHvA
magnetization. The experimental results are compared with
theoretical predictions calculated from the Fermi surface (FS) of
beryllium. The width $h_m$ of the hysteresis loop in the CDS is
measured in a wide temperature and field region. A model for the
hysteresis size is proposed and numerically calculated for the
whole phase diagram.
\end{abstract}
\pacs{75.45.+j, 71.70.Di, 75.60.-d}
\date{\today}
\maketitle

\section{Introduction}

Condon~\cite{Condon1966} predicted the formation of diamagnetic
domains in non-magnetic metals, now known as Condon domains. A
thermodynamic instability arises according to the
Pippard-Shoenberg concept of magnetic
interaction~\cite{Pippard1963,Shoenberg1984} in the de Haas-van
Alphen (dHvA) effect when the amplitude of the oscillatory
magnetization signal becomes large enough, i.e., the
susceptibility
\begin{equation}
\chi =\mu_0\frac{\partial M}{\partial B}>1 \label{equation1},
\end{equation}
where $M$ is the magnetization and $B$ the induction. In this case
the stability condition $\mu_0\partial H/\partial B=1 - \chi >
0$ is not fulfilled for a certain interval of the applied magnetic
field $H$ in each dHvA cycle. For an infinitely long rod-like
sample (demagnetizing factor $n=0$) the system avoids the
instability region by a discontinuous change of the induction $B$
between the two stable states $B_1$ and $B_2$ at a certain
$H=H_c$. Both stable states have the same free energy (see also
Fig.~\ref{Fig9}) and the interval ($B_1,B_2$), containing the
instability, is forbidden.

For a plate-like sample oriented normal to $\mathbf{H}$ ($n=1$) the
boundary condition $B=\mu_0 H$ is required even within the interval
$B_1<\mu_0 H<B_2$, so that the induction $B$ can not change
discontinuously and the state with homogeneous magnetization is
impossible. The plate breaks up into regions of different
magnetization with the inductions $B_1$ and $B_2$. The volume
fractions of the domains is adjusted in a way that the average
induction $\overline{B}=\mu_0 H$ is fulfilled for the whole
sample~\cite{Condon1966}. For a sample with intermediate
demagnetizing factor $0<n<1$ the magnetic field interval $B_1<\mu_0
H<B_2$ with domains decreases proportionally to $n$. Even for
samples of arbitrary shape there is without doubt a non-uniform
Condon domain state (CDS) with the same dia- and paramagnetic phases
$B_1$ and $B_2$. However, the domain configuration is certainly more
complex.

Up to now Condon domains have been observed by different
experimental methods; by NMR~\cite{Condon1968}, $\mu$SR
spectroscopy~\cite{Solt1996,Solt2002} and they were recently
directly observed by Hall probes~\cite{Kramer2005a}. All these
experiments have in common that two distinct inductions $B_1$ and
$B_2$ or an induction splitting $\Delta B=B_2-B_1$ are measured at
a given applied field and temperature.

Equation~\ref{equation1} defines the phase boundary between the
uniform and the CDS which can be calculated using for example the
Lifshitz-Kosevich (LK) formula for the oscillatory dHvA
magnetization resulting from the Landau quantization of the
conduction electrons in a metal. Theoretical calculations of this
boundary exist for several metals~\cite{Gordon1998,Gordon2003}.
However, the above cited measurements yielded only a few points in
the ($H,T$) diagram where Condon domains were actually observed
without a complete determination of the Condon domain phase
diagram. An experimental determination of the CDS phase boundary,
i.e., where $\Delta B$ approaches zero, is difficult and
time-consuming~\cite{Solt1999}. Without doubt, another method for
the experimental determination of the phase boundary is needed to
obtain sufficient data for a comparison with the theoretical
predictions.

Recently, hysteresis was observed in the dHvA effect under the
conditions of the CDS~\cite{Kramer2005b}. Due to the irreversible
magnetization, an extremely nonlinear response to a small
modulation field arises in standard ac susceptibility measurements
upon entering the Condon domain state. The out-of-phase part and
the third harmonic of the pickup voltage rise steeply at the
transition point to the CDS. Moreover, it was shown that the point
($H,T$) where the hysteresis arises is independent of the sample
shape. The threshold character of these quantities allows to
measure a Condon domain phase diagram with high precision and in a
wide temperature and field range. This offers the possibility for
a more detailed comparison with the theoretical calculations.

The CDS boundary of silver has been successfully determined with
this method~\cite{Kramer2010}. The FS parameters of the nearly
spherical FS of silver are well known. Therefore, for silver the
CDS phase diagram in the $(H,T)$ plane can be precisely predicted
using the LK-formula with the Dingle temperature as a
parameter~\cite{Gordon2003}. Good agreement was found with
experimental data of the dHvA oscillation amplitude in the
homogeneous state~\cite{Kramer2005} and for the resulting CDS
phase diagram~\cite{Kramer2010}. This demonstrated that the method
using the nonlinear response for the determination of the CDS
phase boundary is correct.

For beryllium the FS under consideration consists of the well
known electron ''cigars''~\cite{Tripp1969,Shoenberg1984}. The
curvature of the FS at the extremal cross sections is very small
giving rise to a high dHvA amplitude. In addition there are two
close dHvA frequencies of 940~T and 972~T which lead to a beat in
the dHvA amplitude. Due to this frequency beat the CDS phase
diagram is more complex compared to silver. Several models have
been proposed to calculate a CDS phase diagram for beryllium; a
3-dimensional electron gas model using the LK-formula and a purely
2-dimensional electron gas model~\cite{Gordon1998}. However, the
calculations were in contradiction to experimental data obtained
by $\mu$SR. This disagreement required new phase diagram
calculations with a modified LK-formula for the intermediate,
between 2- and 3-dimensional, FS of beryllium taking into account
the real shape of the electron ''cigars''~\cite{Solt2001}. This
model is in good agreement with at that time available $\mu$SR
data. Most recently another theoretical calculation for the phase
diagram was proposed using a different model representation of the
quasi 2-dimensional Fermi surface of
beryllium~\cite{Logoboy2006b}. However, the very few experimental
data available do not allow for a complete test of the recent
calculations over the whole phase
diagram~\cite{Solt2001,Solt2002}.

In this work we determine the experimental Condon domain phase
diagram for beryllium in the whole ($H,T$) plane for $T>1.3$~K
using the appearance of nonlinear response to an ac modulation
field for the detection of the phase
boundary~\cite{Kramer2005b,Kramer2010}. Moreover, the width of the
hysteresis loop in the dHvA effect is measured as function of
temperature and magnetic field in the CDS. Finally, a model for
the origin of the hysteresis is proposed and numerically derived.

\section{Experiment} A standard pickup coil system was used for the
ac measurements of the magnetic susceptibility. The results shown
here were measured on the same rod-like sample as in
Ref.~\onlinecite{Kramer2005b}, of sizes $8\times 2\times 1$~mm$^3$
with the long side being parallel to $[0001]$. The magnetic field
is applied parallel to the long side of the sample. We found from
our measurements a Dingle temperature of $T_D =2.0$~K. The
experiments were carried out at temperatures down to $T=1.3$~K in
a 10~T superconducting coil with a homogeneity of better than
$10^{-5}$ in a sphere with 1~cm diameter. Some experiments were
made in a 16~T coil with a variable temperature insert to measure
temperature dependencies at constant magnetic field. The
modulation frequency was about 160~Hz.

\section{Results}
\begin{figure}[tb]
\begin{center}
\includegraphics*[width=\linewidth, bb=12 4 200 242]{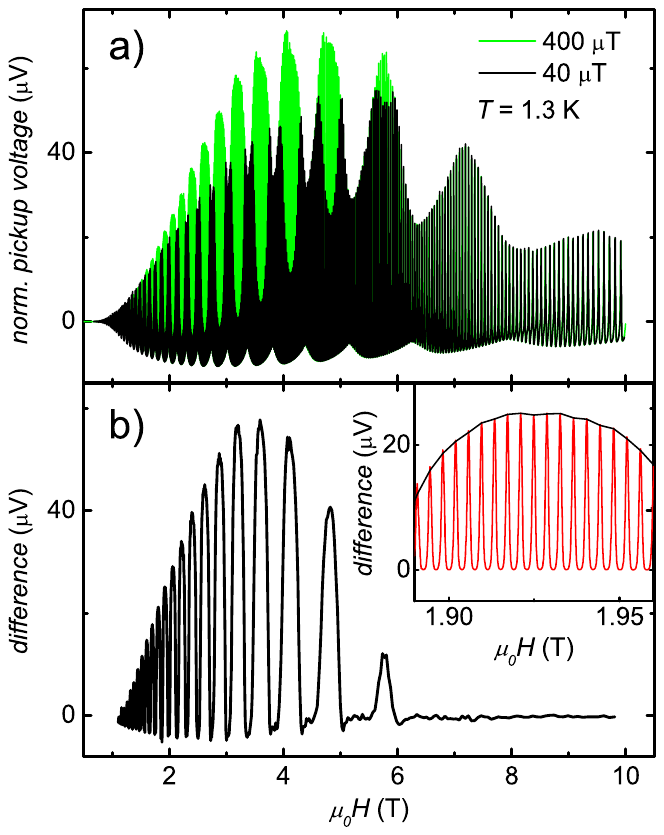}
\caption{(a)~Pickup voltage normalized on the modulation level for
high (400~$\mu$T) and low (40~$\mu$T) modulation level at 1.3~K. Due to
hysteresis in the CDS the pickup voltage decreases if the
modulation level is of the order or smaller than the hysteresis
loop width. (b)~Envelope of the difference between both curves of
Fig.~\ref{Fig1}(a) showing the field regions where Condon domains
exist for non-zero difference. The inset shows an expanded view of
the difference signal with the appearance of hysteresis in a part
of each dHvA period. \label{Fig1}}
\end{center}
\end{figure}
The phase transition point to the CDS can be determined by several
methods~\cite{Kramer2010} which are all based on the appearance of
hysteresis in the dHvA effect~\cite{Kramer2005b}. Figure~\ref{Fig1}
shows the pickup voltage normalized on the modulation level for low
and high modulation amplitude in a large magnetic field range at
1.3~K. Due to the hysteresis in the CDS the response to an ac
modulation becomes extremely nonlinear and the first harmonic
amplitude of the pickup voltage normalized on the modulation
amplitude, usually corresponding to the susceptibility $\chi$,
decreases strongly at the paramagnetic part ($\chi>0$) of every dHvA period. The amplitude
damping is observed if the modulation level is of the order or
smaller than the width of the hysteresis loop. In absence of Condon
domains the normalized pickup voltage is independent of modulation
level. Therefore, the substraction of two curves, one measured with
high and the other with low modulation level, reveals the magnetic
field ranges where Condon domains exist. In other words domains
exist if the difference is greater than zero. Figure~\ref{Fig1}(b)
shows the envelope of this function at $T=1.3$~K. The inset presents
the detailed difference of both curves in a small region, which
corresponds to a cut of the CDS phase diagram at $T=1.3$~K. We see
in Fig.~\ref{Fig1}(b) that there is no difference between the
normalized pickup voltages for magnetic fields exceeding 6~T which
implies that Condon domains disappear for fields higher than 6~T at
1.3~K.

For the above described method to determine the CDS phase diagram,
two field sweeps must be measured for each temperature. In order
to detect even very small hysteresis the low modulation level must
be as small as possible. Therefore, it is difficult to detect by
this method the existence of Condon-domains at field regions where
the hysteresis loop width is small.

\begin{figure}[tb]
\begin{center}
       \includegraphics*[width=\linewidth, bb=2 0 270 197]{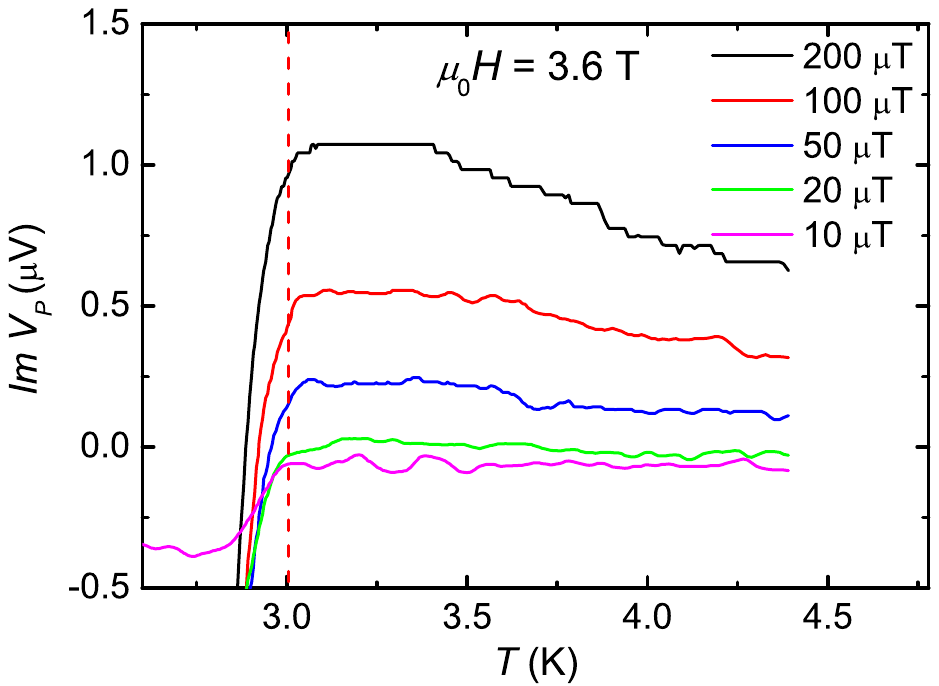}
       \caption{
Temperature dependence of the imaginary part of the pickup voltage
for several modulation amplitudes at the paramagnetic part of a
dHvA oscillation at the beat antinode at 3.6~T. The dashed line
indicates the critical temperature $T_c=3.0$~K where the Condon
domain phase is entered when lowering the temperature.
\label{Fig2}}
\end{center}
\end{figure}
It was shown that the out-of-phase part and the third harmonic of the pickup voltage
appear with threshold character whenever there is small hysteresis
in the dHvA magnetization~\cite{Kramer2005b}. A measurement of one
of these quantities offers therefore a simple alternative way to
determine with high precision the phase boundary of the CDS.
Figure~\ref{Fig2} shows temperature dependencies of the
out-of-phase part of the pickup voltage for a large modulation
level range at the beat antinode of the dHvA oscillations at a
maximum of $\chi$ near 3.6~T. At the critical temperature
$T_c=3.0$~K the out-of-phase signal drops down rapidly upon
lowering the temperature. This indicates a sudden phase shift of
the ac response with respect to the modulation signal. The phase
shift is caused by the emerging hysteresis in the dHvA
magnetization~\cite{Kramer2005b}. We find the same $T_c$,
indicated by the dashed line in Fig.~\ref{Fig2}, for all
modulation levels showing that the determination of the CDS phase
boundary is independent of the used modulation level.

In the following we will use a modulation amplitude of 40~$\mu$T. This
value is sufficiently small to insure that the dHvA period is
always much bigger than the modulation amplitude $h$ even at low
magnetic fields (at $\mu_0 H=1$~T the dHvA period of beryllium is
about 1~mT). On the other hand the ac response is still easily
detectable. We note that if the modulation amplitude is of the
order of the dHvA period the imaginary part and the third harmonic
of the pick-up signal show up even in the absence of
hysteresis~\cite{Shoenberg1984}.

\begin{figure}[tb]
\begin{center}
       \includegraphics*[width=\linewidth, bb=2 1 260 201]{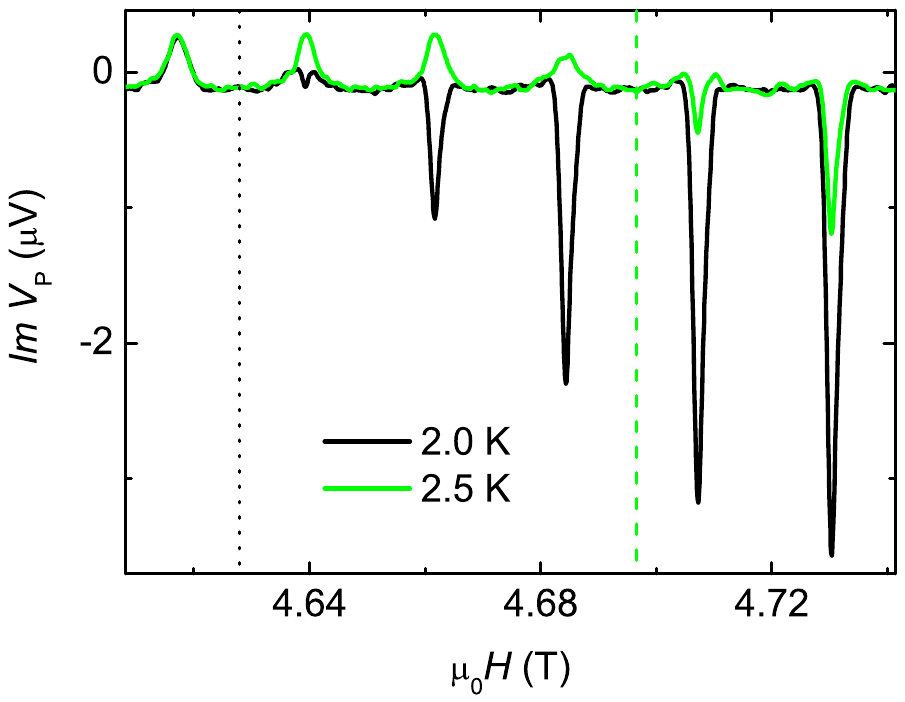}
       \caption{
Field dependence of the imaginary part of the pickup voltage
measured at 2.0~K and 2.5~K with 40~$\mu$T modulation amplitude. The
steeply increasing negative amplitude is caused by the hysteresis
in the CDS. The vertical dotted lines indicate the transition
field to the CDS for 2.0 and 2.5~K. \label{Fig3}}
\end{center}
\end{figure}

Figure~\ref{Fig3} shows magnetic field dependencies of the
imaginary part of the pickup voltage measured at 2.0~K and 2.5~K.
The phase of the lock-in amplifier is adjusted such that the
signal due to the sample susceptibility is mainly in-phase. dHvA
oscillations of small amplitude similar to the waveform in the
inset of Fig.~\ref{Fig1} are visible in the out-of-phase signal
with an amplitude bigger than previously observed in
Ref.~\onlinecite{Kramer2005b}, due to the increased modulation
frequency of 160~Hz compared to 21~Hz in
Ref.~\onlinecite{Kramer2005b}. The higher eddy currents explain
the appearance of an out-of-phase signal for an homogeneous
magnetization. However, Fig.~\ref{Fig3} shows threshold character
in the arising of negative peaks at magnetic fields where
hysteresis occurs indicating the transition to the CDS around the
negative peaks in each dHvA period. We see in Fig.~\ref{Fig3} that
the negative peaks appear at 2.0~K at lower fields than at 2.5~K.
At temperatures above 3.0~K all negative peaks disappear and only
the small dHvA oscillations due to eddy-current effects remain.

The amplitude of the negative peaks depends on the modulation
level, the hysteresis loop width at the particular magnetic field,
and on the amplitude of the in-phase part of the pickup voltage,
i.e. the susceptibility. Even though the peak amplitude seems to
be correlated with the Condon domain phase diagram being stronger
further away from the phase-diagram boundary, we extract from
these data only the magnetic field values for which the negative
peaks appear for each temperature in order to construct the phase
diagram in the next section. The negative peaks arise with
threshold character and Fig.~\ref{Fig3} shows that the CDS phase
boundary can be determined with a precision of about one dHvA
period.

It was reported in $\mu$SR
studies~\cite{Solt1999,Solt2001,Solt2002} that Condon domains
occur also at the beat nodes of the dHvA oscillations around 2.0~T
and 2.7~T for 0.5~K and 0.8~K, respectively. However, there are
only a few temperature dependencies of the induction splitting
available from $\mu$SR measurements. In other words, the reported
temperatures do not represent necessarily the CDS phase boundary
for these fields.
\begin{figure}[tb]
\begin{center}
       \includegraphics*[width=\linewidth, bb=2 0 297 193]{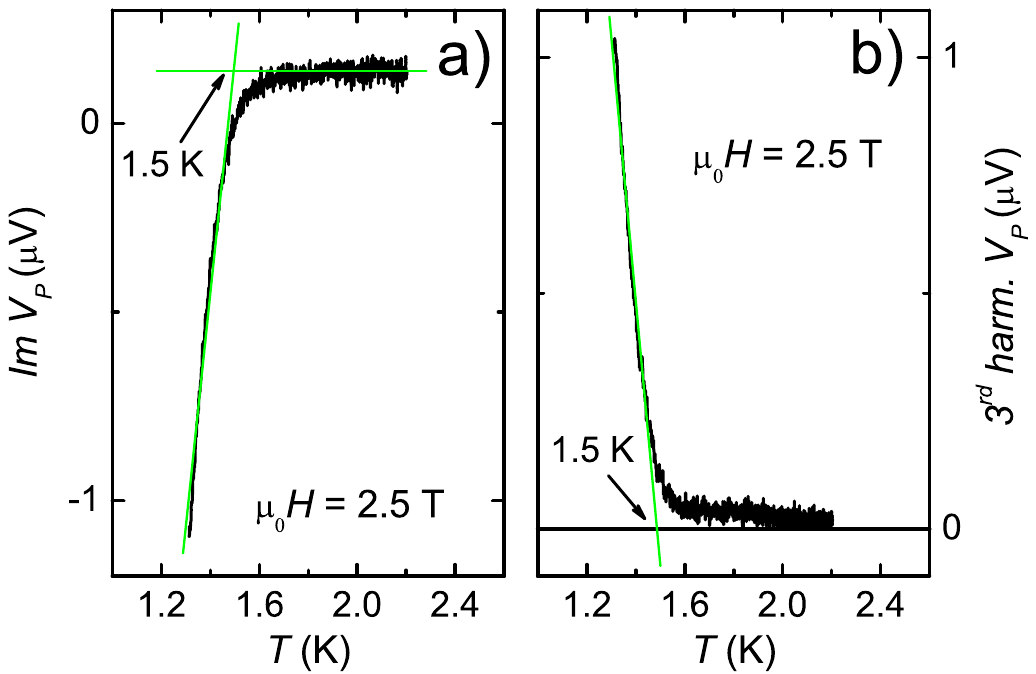}
       \caption{Temperature dependence of the imaginary part~(a)
       and the third harmonic~(b) of the pickup voltage measured in the
       paramagnetic part of a dHvA oscillation at
       the beat node for 2.5~T with 40~$\mu$T modulation amplitude.
       \label{Fig4}}
        \end{center}
\end{figure}
Figure~\ref{Fig4} shows the temperature dependence of the
out-of-phase part~(a) and the third harmonic~(b) of the pickup
voltage at the beat node at 2.5~T. A sharp transition at 1.5~K is
visible in both traces which indicates that hysteresis arises at
this temperature. This means that Condon domains appear indeed at
this beat node and the CDS phase boundary is at 1.5~K for 2.5~T.

\section{Phase diagram}
\begin{figure}[tp]
\begin{center}
      \includegraphics*[width=\linewidth, bb=2 1 256 200]{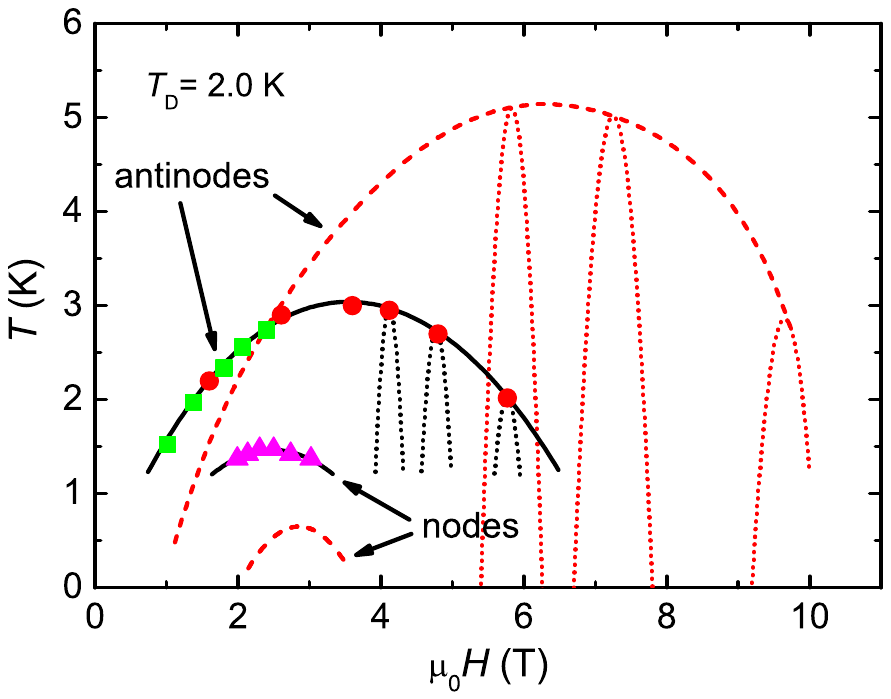}
       \caption{
Phase diagram in the ($H,T$) plane for beryllium. Scatter points
indicate the position of the phase boundary at beat nodes
and antinodes determined by temperature sweeps (circles and
triangles) like in Figs.~\ref{Fig2} and~\ref{Fig4} and by field
sweeps (squares) like in Fig.~\ref{Fig3}. The solid lines are
guiding lines of the phase boundary for beat nodes and antinodes,
respectively. The dashed lines show for comparison the recent
theoretical calculation of Ref.~\onlinecite{Solt2001,Solt2002}.
The dotted lines indicate the envelope of the beating substructure
of the phase diagram shown in detail in
Fig~\ref{Fig6}.\label{Fig5}}
\end{center}
\end{figure}
\begin{figure}[tb]
\begin{center}
\includegraphics*[width=\linewidth, bb=2 1 239 197]{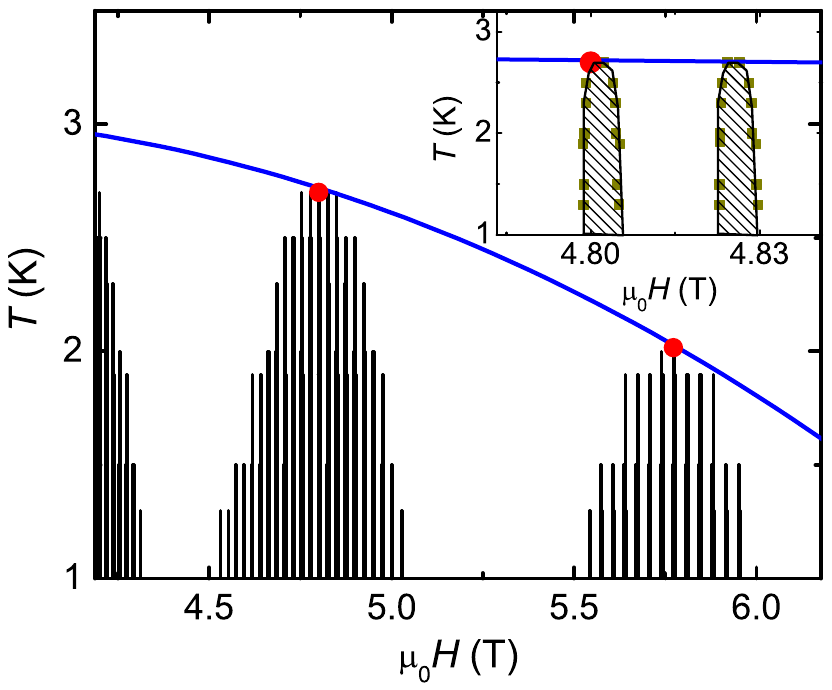}
\caption{Expanded view of Fig.~\ref{Fig5} around the beat maximum
at 4.8~T showing the beating substructure of the phase diagram.
The slight steps in the envelope of the height of the needle-like
stripes results from the limited number of measured temperatures.
The expanded view in the inset shows the detailed phase diagram in
two successive dHvA periods obtained by field sweeps at constant
$T$. The larger round circles are obtained by $T$-sweeps at
constant $H$. \label{Fig6}}
\end{center}
\end{figure}

We have seen that the CDS phase boundary can be determined with
high precision using nonlinear response measurements. Due to the
hysteresis the out-of-phase signal of the pickup voltage drops
sharply. This was measured either at a fixed magnetic field like
in Fig.~\ref{Fig2} and~\ref{Fig4} or at fixed temperature as
function of magnetic field like in Fig.~\ref{Fig3}. All data is
compiled to obtain a complete Condon domain phase diagram in
Fig.~\ref{Fig5}. The solid lines in Fig.~\ref{Fig5} are
extrapolated guiding lines to the ($H,T$)-values obtained for the
beat antinodes and nodes, respectively. These lines are the
envelopes of a substructure consisting of a beating pattern of
sharp needle-like domain regions as shown in Fig.~\ref{Fig6} where
the inset reveals the Condon-domain regions in two successive dHvA
periods. For magnetic fields in between these needle-like regions
the sample is in the homogeneous state. We see in Fig.~\ref{Fig6}
that Condon domains appear first for magnetic fields around a beat
antinode where the dHvA amplitude is higher. When cooling down the
CDS field range extends gradually around the antinodes.

The condition (see Eq.~1) that a Condon domain state occurs in
a dHvA period is independent of the demagnetization factor. We have
found in a test on a plate-like sample with the same Dingle
temperature that the obtained phase diagram is indeed independent of
the sample shape. However, this only holds for the envelope of the
phase diagram (solid lines in Fig.~\ref{Fig5}). The substructure
depends on the sample shape as the needle-like regions (inset of
Fig.~\ref{Fig6}) are much broader for a plate-like sample. The
reason for this is that the field range within a dHvA period where
domains arise scales with $n$ and is therefore more extended in a
plate-like sample~\cite{Shoenberg1984}. In other words, the envelope
of the phase diagram in the ($H,T$)-plane is independent of the
demagnetization factor, but not the detailed field region within a
single dHvA period.

The phase diagram in Fig.~\ref{Fig5} agrees with all reported
$\mu$SR data~\cite{Solt1996,Solt1999,Solt2001,Solt2003}. In
particular, the observed induction splitting disappeared at the beat
antinode near 2.6~T for temperatures higher than 3.0~K. We examined
the same beat maximum and found a critical temperature of 2.9~K for
our sample.

In Fig.~\ref{Fig5} the experimental phase diagram is compared with
the calculations made with the modified LK-formula in
Ref.~\onlinecite{Solt2001,Solt2002}. The overall shape of the
calculated antinode and node envelope curves is similar to the
experimental result. However, there is a clear discrepancy
between the predicted temperature and magnetic field ranges of the CDS and the ones we find experimentally. We observe an upper critical field of about 8~T (extrapolation of
the guide to the eye for the antinodes in Fig.~\ref{Fig5} to $T=0$)
above which domains disappear for all temperatures for our sample
with $T_D=2.0$~K. We see in addition that Condon domains
continue to exist at higher temperatures down to lower fields
compared to the phase-diagram calculation and that domains exist at the beat nodes up to higher temperatures than predicted. A reason for the
discrepancy in the temperature-field values of the phase-diagram
boundary can be possibly related to the strong magnetostriction
effects in beryllium (see discussion below).

\section{Hysteresis loop size}

\begin{figure}[tb]
\begin{center}
       \includegraphics*[width=\linewidth, bb=35 35 562 430]{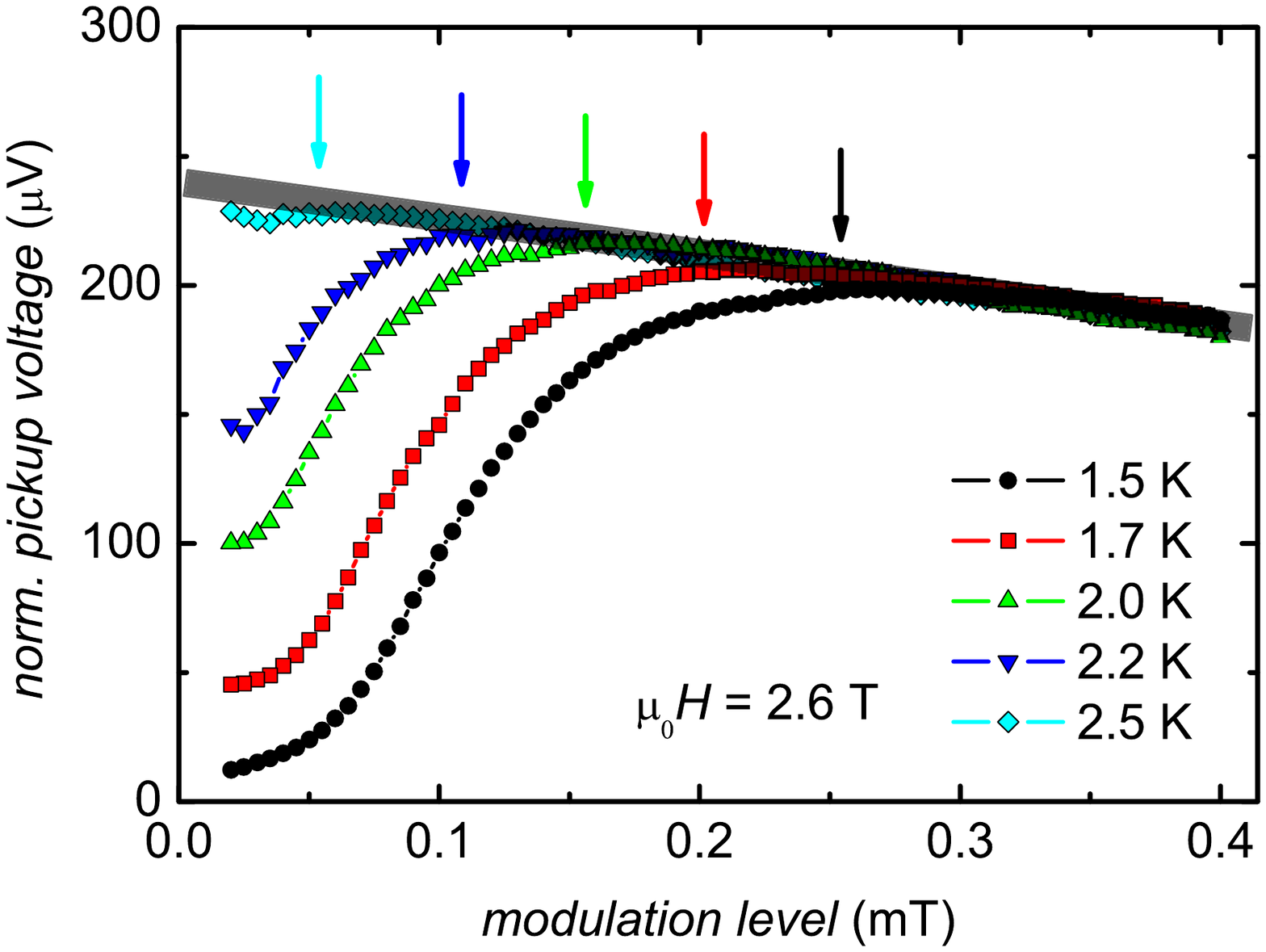}
        \caption{
Normalized pickup voltage as function of modulation level for
several temperatures in the center of the dHvA oscillation at 2.6~T
corresponding to the center of a needle-like stripe like shown in
Fig.~\ref{Fig6}. The arrows indicate the chosen hysteresis
sizes $h_m$.\label{Fig7}}
       \includegraphics*[width=\linewidth, bb=49 16 254 203]{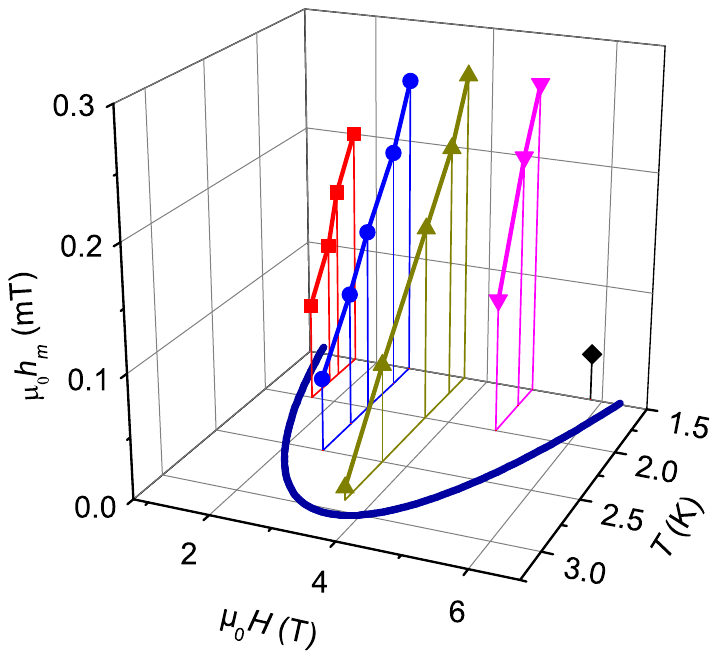}
       \caption{
Hysteresis amplitude $h_m$ as function of magnetic field and
temperature (data points), determined like in Fig.~\ref{Fig7} as a
measure for the hysteresis width. The parabola indicates the region of
the CDS phase diagram in the ($H,T$) plane (solid line of the antinodes in Fig.~\ref{Fig5}). \label{Fig8}}
\end{center}
\end{figure}

Another interesting question is the dependence of the hysteresis
loop size $h_m$ from temperature and magnetic field. At the CDS
phase boundary $h_m$ vanishes. However, the temperature and field
dependence of $h_m$ in the CDS might give information about the
nature of the irreversible magnetization. We suppose that the
hysteresis itself and its size are mainly caused by
irreversible domain wall motion or by nucleation of new domain
fractions. In the following we will measure the temperature and
field dependence within the phase diagram.

As it was shown earlier~\cite{Kramer2005b} $h_m$ can hardly be
measured directly by Hall probes because of its small magnitude.
However, $h_m$ can be indirectly determined by analyzing the
response characteristic to an ac modulation field. As shown above,
the normalized pickup voltage decreases strongly if the modulation
amplitude decreases below the hysteresis size (see Fig.~\ref{Fig7}).
All measurements were made in the center of a dHvA period, i.e., in
the center of a needle-like stripe of the phase diagram (see
Fig.~\ref{Fig6}). From Fig.~\ref{Fig7} the order of magnitude of
$h_m$ can be estimated. However, it is not obvious which modulation
level corresponds actually to the real $h_m$ which would be observed
with Hall probes. A comparison with direct measurements by Hall
probes under the same conditions yields good agreement if we chose
for $h_m$ the onset of the decrease in the normalized pick-up voltage indicated by arrows in Fig.~\ref{Fig7}.
All data points obtained in this way are presented in Fig.~\ref{Fig8}. We see a more or less linear temperature dependence
of $h_m$ far enough from the phase boundary. Moreover, $h_m$ is
practically independent of $H$ at the lowest measuring temperature
far from the phase boundary.

\section{Model for hysteresis loop size}

Hysteresis in the CDS is certainly due to irreversible domain wall
motion or rearrangement processes of the respective domain volume
fractions upon field variation. In the following we analyze the
shape and amplitude of the energy barrier in the domain wall
between the phases with the inductions $B_1$ and $B_2$.

\begin{figure}[tb]
\begin{center}
       \includegraphics*[width=0.7\linewidth, bb=2 2 232 192]{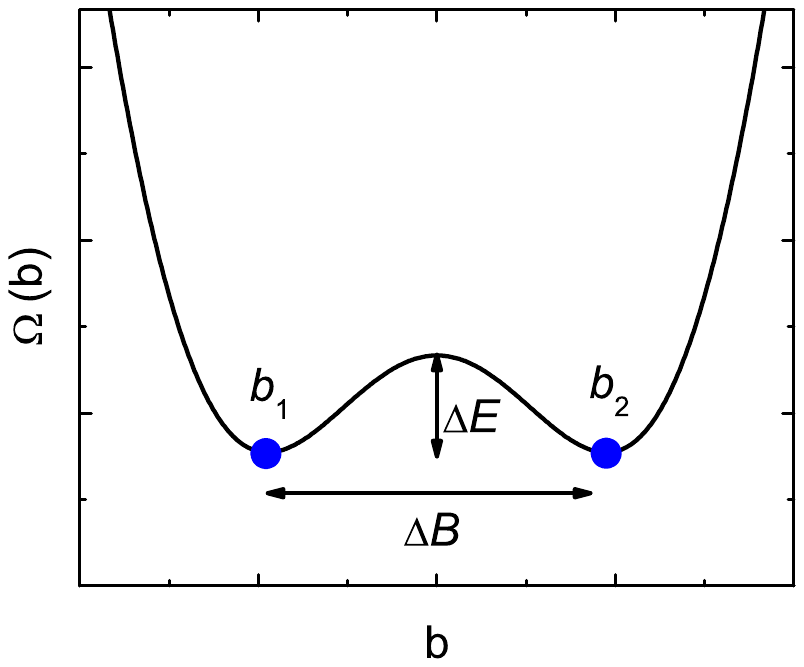}
        \caption{
Schematic representation of the Lifshitz-Kosevich potential in the
CDS showing two minima corresponding to the domain states with the
inductions $B_1=\mu_0H+b_1$ and $B_2=\mu_0H+b_2$ and a potential barrier $\Delta E$ in the domain
wall between the domains.
\label{Fig9}}
\end{center}
\end{figure}

We can write the potential $\Omega$ as the sum of the contribution
of the magneto-quantum oscillations (taking only the first harmonic of
the LK-formula) and the magneto-static energy
\begin{equation}
    \Omega=a(H,T) \cos\left(2\pi \frac{F}{B}\right) +
    \frac{1}{2\mu_0}(B-\mu_0H)^2. \label{equation2}
\end{equation}
with the dHvA frequency $F$ and the oscillation amplitude $a(H,T)$ given by the LK-formula~\cite{Shoenberg1984}.
For inductions $B$ close to the magnetic field $H$ we can develop Eq.~\ref{equation2} setting $B=\mu_0H+b$
\begin{equation}
    \Omega=a(H,T) \cos\left(\frac{2\pi}{p}b\right) +
    \frac{1}{2\mu_0}b^2 \label{equation3}
\end{equation}
with the dHvA period $p=(\mu_0H)^2/F$. If the
amplitude $a(H,T)$ is big enough, which corresponds to the condition in Eq.~\ref{equation1}, then $\Omega$
has two minima at $b_1$ and $b_2$. Figure~\ref{Fig9} shows
schematically the potential under this condition.

The states between $B_1$ and $B_2$ have extra energy and are not stable.
However, in the domain wall the induction $B$ has to cross all
values between $B_1$ and $B_2$. Therefore, there is an energy
barrier (see Figure~\ref{Fig9}) separating the states with these
inductions whose amplitude $\Delta E$ can be calculated. We would expect that $\Delta E$ scales with the hysteresis width because this
energy barrier must be overcome when the domain distribution
changes under variation of the applied field.

\begin{figure}[tb]
\begin{center}
       \includegraphics*[width=\linewidth, bb=97 272 485 556]{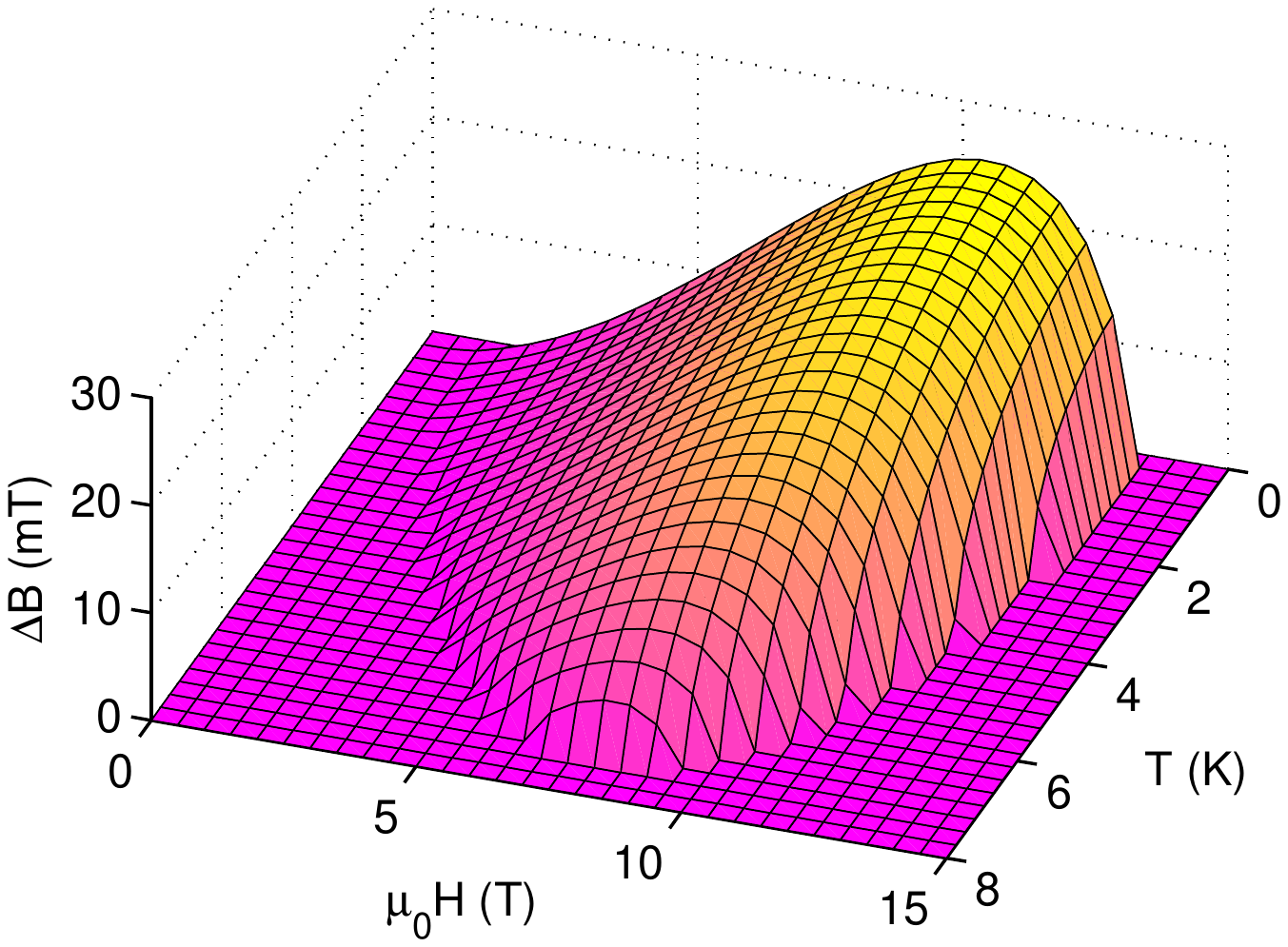}
        \caption{
Calculation of the induction difference $\Delta B$ between the domains as a function of $H$ and
$T$ for a Dingle temperature $T_D=2.0$~K.\label{Fig10}}
\end{center}
\end{figure}

First, the inductions values $B_1$ and $B_2$ of the domains are
found by minimization of the free energy
\begin{eqnarray}
    \frac{\partial \Omega}{ \partial b}=0 \mbox{, or} \quad
    a(H,T) \frac{2\pi}{p}\sin\left(\frac{2\pi}{p}b_0\right)= \frac{b_0}{\mu_0}
\end{eqnarray}
where
\begin{equation}
b_0=\frac{b_2 - b_1}{2}=\frac{\Delta B}{2}.
\end{equation}
The induction difference $\Delta B$ between the domains is
calculated as a function of $H$ and $T$ in Fig.~\ref{Fig10}. Here, and in the following calculations we use the 'cylinder' model of Ref.~\onlinecite{Solt2001} which approximates the cigar-like Fermi surface shape of beryllium with a cylinder to determine $\chi(H,T)$ and $a(H,T)$. This idealized model gives a reasonable upper limit for the phase boundary of the domains and should be sufficient to get an idea of the overall behavior of $\Delta E$ as a function of $H$ and $T$.

Once $b_1$ and $b_2$ are known we can calculate the amplitude of the energy barrier $\Delta E$ using the above formulas
\begin{eqnarray}
    \Delta E & = & \Omega(0)-\Omega(b_1) \nonumber \\
    & = & a(H,T) - a(H,T) \cos\left(\frac{2\pi}{p}b_1\right) - \frac{b_1^2}{2\mu_0}.
\end{eqnarray}

\begin{figure}[tb]
\begin{center}
       \includegraphics*[width=\linewidth, bb=87 262 507 578]{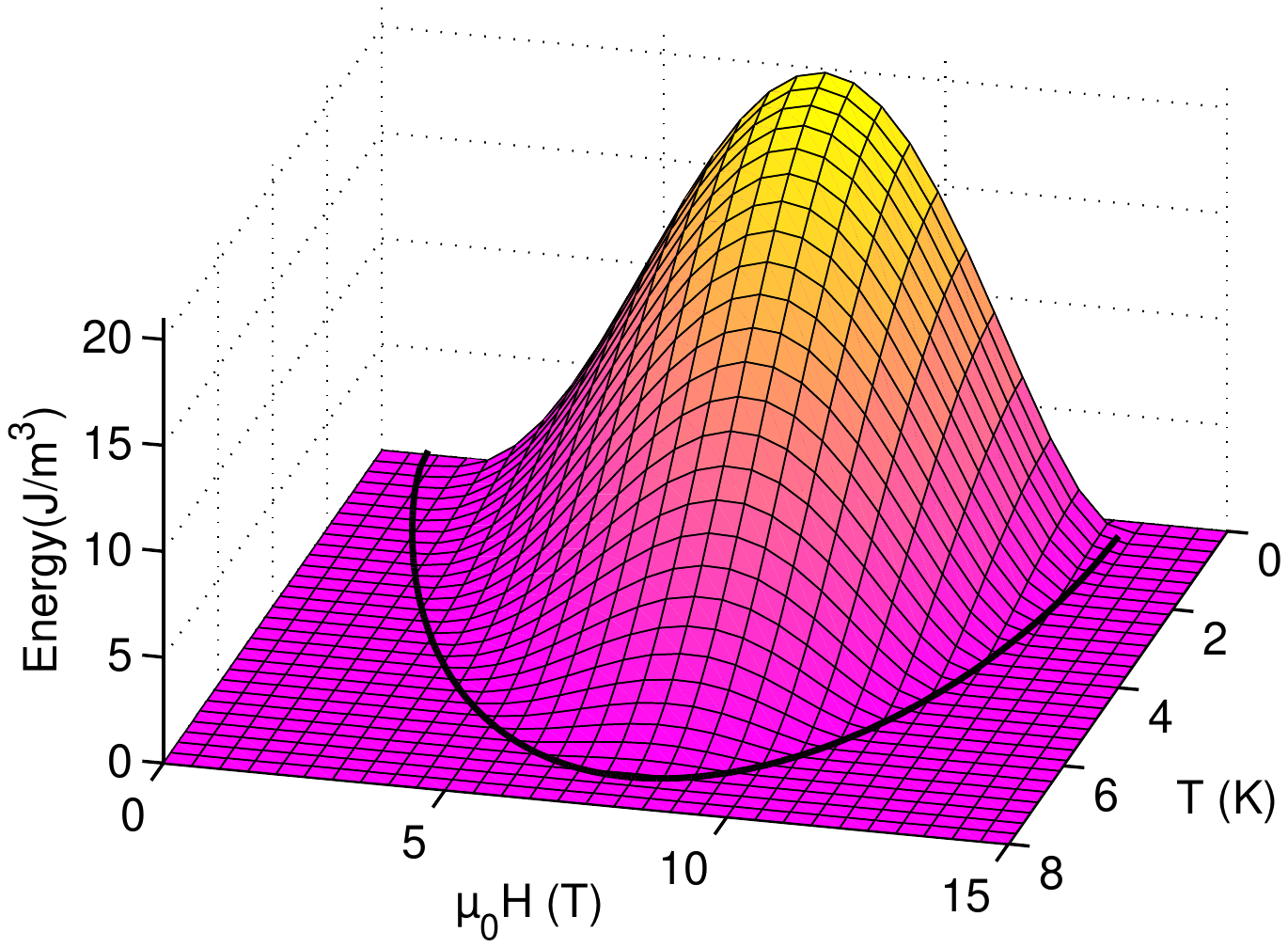}
        \caption{
Numerical calculation of the energy barrier $\Delta E$ between the domains with
the respective inductions $B_1$ and $B_2$ as function of $H$ and
$T$ for a Dingle temperature $T_D=2.0$~K. The solid line indicates the Condon domain phase boundary using the 'cylinder' model of Ref.~\onlinecite{Solt2001}. \label{Fig11}}
\end{center}
\end{figure}

A simpler expression for $\Delta E$ can be given in good approximation taking into account that the shape of $\Omega(b)$ between $b_1$ and $b_2$ is very similar to the cosine-function
\begin{equation}
 \Omega \approx \frac{\Delta E}{2}\left[\cos\left(\frac{\pi}{b_0}b\right)+1\right].
\end{equation}
Taking the second derivative of this function with respect to $b$ and taking into account that the curvature of $\Omega$ is $1/\mu_0-\chi$ at $b=0$, (here $\chi$ is positive and $\mu_0\chi>1$, we find the following expression
\begin{equation}
\Delta E = \frac{1}{2 \pi^2 \mu_0}\left(\mu_0 \chi - 1\right) \Delta B^2.
\end{equation}
This expression for $\Delta E$ is similar to the domain wall surface energies calculated earlier~\cite{Privorotskii1967,Abrikosov1988}.
Figure~\ref{Fig11} shows the numerical calculation of $\Delta E$ as a function of
temperature and magnetic field.

We see a qualitative agreement between the calculated
behavior of $\Delta E$ with the observed width of the hysteresis
loop in Fig.~\ref{Fig8}. Both show a maximum at approximately the middle of the magnetic field range of the CDS phase boundary at lowest
temperature. There is also good agreement with a recent theoretical
calculation where the Rayleigh model was used for the hysteresis~\cite{Logoboy2006a}. However, we notice that the energy barrier model
shows a clear difference from our data near the phase boundary. $h_m$ varies rather steeply when approaching the phase boundary whereas $\Delta E$ increases only gradually at the phase
boundary. One can say that our model describes well the
Condon domain hysteresis far from the phase boundary where
the irreversibility results mainly from domain wall motion. We
suggest therefore that close to the phase boundary a mayor
contribution to the hysteresis is due to the nucleation of new phase
fragments in tubular form. The region where the tubular structure undergoes a transition to a
laminar one with stripes could be very narrow compared to the intermediate state of type-I superconductors~\cite{Egorov2001}.

\section{Discussion}
We can summarize two main differences in the behavior between
beryllium and silver. First, as it was shown recently by
using local Hall probes~\cite{Kramer2005a}, Condon domains do not
emerge to the sample surface in beryllium, appearing only inside the
bulk. For silver the measured inductions values of the respective
domains are practically the same inside the bulk and on the surface
indicating that domains emerge completely to the surface. Second,
as it is shown here, the experimental CDS boundary lies inside
the closest model and the discrepancy increases for higher
magnetic fields unlike to silver where at least up to 30~T good
agreement was observed with theoretical
calculations~\cite{Kramer2010}. We propose that the reason for
both findings could lay in the fact that the dHvA effect is always
accompanied by magnetostriction oscillations. In the particular case
of the CDS this means that domain formation gives rise to different,
actually opposite, deformations in the neighboring
domains~\cite{Egorov2002,Egorov2004}.

This deformation varies across the domain walls between
neighboring domains and requires extra elastic energy in the domain
walls and on the surface. Moreover, the magnetostriction amplitude
increases with magnetic field and the amplitude is actually very big
especially for beryllium. Beryllium has in comparison to silver a
much higher Young modulus and the deformation under the dHvA effect
is anisotropic. This idea can qualitatively explain the
discrepancy in the phase diagram between theory and experiment at
high magnetic fields. Nevertheless, the discrepancy remains for low magnetic
fields and for the nodes where the experimentally obtained
boundaries are above the calculated ones. It would be of interest to
include the influence of magnetostriction on the calculated phase
boundaries.

\section{Conclusion} We have measured a complete Condon domain
phase diagram for beryllium at temperatures down to 1.3~K and
magnetic fields from 1~T up to 10~T. The method based on the
detection of the nonlinear response to an ac modulation field
provided also information about the substructure of the phase
diagram due to the dHvA frequency beat in beryllium. The
measurements agree with all data obtained by $\mu$SR. Moreover, we
have checked that the obtained phase diagram is independent on the
sample shape. The method can be easily applied to samples with
other Dingle temperatures and other metals.

The hysteresis loop size was measured in a wide region of the CDS
phase diagram. In the middle of the phase diagram, far enough from
the boundary, the hysteresis loop width increases linear with
decreasing temperature and it is almost constant with magnetic
field. Finally, a model for the origin of the hysteresis is
proposed. The induction difference between the different domains is
numerically derived and the height of the energy barrier separating
these two states of induction is calculated. We found that the
calculated energy barrier scales well with the observed hysteresis
loop width besides in the region close to the phase boundary. We suggest that close to the phase
boundary the domain wall motion is not the only reason for the observed
hysteresis and that the process of filamentary nucleation of the
newly created phase must be taken into account.

\begin{acknowledgments}
We are grateful to N.~Logoboy, I.~Sheikin and V.P.~Mineev for
fruitful discussions.
\end{acknowledgments}

\bibliography{REF_CondonDomain2}

\end{document}